\documentstyle[aps,prl,epsf,multicol]{revtex}
\begin{document}
\draft
\title{
Effects of carrier concentration on the superfluid density of 
high-T$_c$ cuprates }  
\author{C. Panagopoulos$^1$,  B.D. Rainford$^2$,  J.R. Cooper$^1$, W. Lo$^1$, 
J.L. Tallon$^3$, J.W. Loram$^1$, J. Betouras$^4$,  Y.S. Wang$^5$  
and  C.W. Chu$^5$}
\address{
$^1$IRC in Superconductivity, University of Cambridge, Cambridge CB3 0HE, 
United Kingdom \\
$^2$Department of Physics and Astronomy, University of Southampton, 
Southampton SO17 1BJ, United Kingdom \\
$^3$New Zealand Institute for Industrial Research, P.O. Box 31310, Lower Hutt, 
New Zealand \\
$^4$Department of Physics, Theoretical Physics, University of Oxford, 
Oxford, OX1 3NP, United Kingdom \\
$^5$Department of Physics and Texas Centre for Superconductivity at 
the University of Houston, Houston, Texas 77204-5932}
\date{today}
\maketitle

\begin{abstract}

The absolute values and temperature, T, dependence of the
in-plane magnetic penetration depth, $\lambda_{ab}$, of 
La$_{2-x}$Sr$_x$CuO$_4$ and HgBa$_2$CuO$_{4+\delta }$ have been measured as
a function of carrier concentration. We find that the superfluid
density, 
$\rho_s$, changes substantially and systematically with doping. 
The values of $\rho_{s}(0)$ are closely linked to the available low
energy spectral weight as determined by the electronic entropy just
above T$_c$
 and the initial slope of $\rho_{s}(T)/\rho_{s}(0)$ 
increases rapidly with carrier concentration. 
The results are discussed in the context of a possible relationship
between 
$\rho_s$ and the normal-state (or pseudo) energy gap.

\end{abstract}

\begin{multicols}{2}
Superconductivity arises from the binding of electrons into
Cooper pairs thereby forming a superfluid with a superconducting
energy gap, 
$\Delta$, in the single-particle excitation spectrum. In
high-temperature 
superconductors (HTS) $\Delta$ has essentially $d_{x^2-y^2}$ symmetry
in 
k-space with $\Delta_k = \Delta_0 cos(2\phi)$ [1], where 
$\phi = arctan(k_y / k_x)$ and $\Delta_{0}$ is the superconducting 
gap amplitude which will in general be $\phi$ dependent. 
Changes in carrier concentration affect the superconducting [2-8] and 
normal state (NS) [4-7] properties of HTS and there is evidence [3-7]
that 
in addition to $\Delta_k$ there is a normal state (or pseudo) gap,
$\Delta_N$,
 in the NS energy excitation spectrum in under- and optimally doped
samples 
which increases with decreasing doping. The maximum gap amplitude
shows 
little variation with underdoping even though T$_c$ is reduced [3-8], 
in disagreement with the standard mean-field Bardeen-Cooper-Schrieffer
(BCS) theory. This unusual behaviour is probably linked to the
presence 
of $\Delta_N$ [7]. However, fundamental problems such as the origin of
$\Delta_N$ and its possible effect on the superfluid density,
$\rho_s$, have not been clearly resolved experimentally or theoretically. 

The physical quantity most directly associated with $\rho_s$
is the magnetic penetration depth $\lambda$. 
Appropriate systems to investigate $\rho_s$ as a function of doping
are 
La$_{2-x}$Sr$_x$CuO$_4$ and HgBa$_2$CuO$_{4+\delta }$. 
Both have a simple crystal structure with one CuO$_2$ plane per unit
cell, 
can have their carrier concentration controlled, 
and there is experimental evidence suggesting the presence of
$\Delta_N$ 
which closes with doping [3-5,9]. Here we report in-plane penetration
depth, 
$\lambda_{ab}$, measurements for high-quality La$_{2-x}$Sr$_x$CuO$_4$
(LSCO) with x = 0.10, 0.15, 0.20, 0.22, 0.24 measured by the 
ac-susceptibility ($acs$) and muon spin relaxation ($\mu$SR)
techniques and 
for HgBa$_2$CuO$_{4+\delta }$ (Hg-1201) with $\delta$ = 0.10, 0.37 measured only by
$\mu$SR. We find systematic changes in $\rho_s$ with carrier 
concentration and a correlation with $\Delta_N$.
	
Single-phase polycrystalline samples of LSCO were prepared in 
Cambridge using solid-state reaction procedures. 
No other phases were detected by powder x-ray diffraction and the 
phase purity is thought to be better than 1\%. Lattice parameters were 
in good agreement with published work [10]. High field magnetic 
susceptibility measurements showed no signatures of excess
paramagnetic 
centres. The measured T$_c$'s are 30, 37.7, 36, 27.5 and 20.3 K for x
= 0.10, 0.15, 0.20, 0.22 and 0.24, respectively.  These values are
also in very good agreement with previous measurements [10]. 
$\mu$SR experiments as a function of T were performed on the same
powders for x = 0.10 and 0.15. Although unoriented powders can be used
to determine $\lambda_{ab}$ by $\mu$SR [2], the $acs$ technique requires
the powders to be magnetically aligned [11]. To eliminate grain 
agglomerates, powders were ball-milled in ethanol and dried after
adding a defloculant. Scanning electron microscopy confirmed the
absence of grain boundaries and showed that the average grain diameter
was $\sim 5 \mu m$. The powders were mixed with a 5 min curing epoxy 
and aligned in a static field of 12T at room temperature. 
Debye-Scherrer x-ray scans showed that $\sim 90 \%$ of the grains were 
aligned within $\sim {2.0}^{o}$. Low-field susceptibility measurements
were performed at an $ac$-field H$_{ac}$ = 1 G rms (parallel to the
c-axis) and a frequency f = 333 Hz down to 1.2K. Details of the
application of London's model for deriving $\lambda$ from the measured
susceptibility can be found in an earlier publication [11]. 
Transverse-field-cooled $\mu$SR experiments were performed at 400 Gauss in
the ISIS, Rutherford-Appleton Laboratory. 
The field produced a flux-line lattice whose field distribution was
probed by muons. The depolarisation rate, $\sigma (T)$, of the initial
muon spin is proportional to ${\lambda_{ab}}^{-2}(T)$ [2,12]. 
Checks were made to ensure that the values of $\lambda_{ab}$ obtained 
were independent of the applied field. The Hg-1201 [$\delta$ = 0.10 
(T$_c$ = 60K) and 0.37 (T$_c$ = 35K)] samples were prepared in Houston
by the controlled solid-vapour reaction technique [13].

The values of $\lambda_{ab}(0)$ derived from the $acs$ data
for LSCO are 0.28, 0.26, 0.197, 0.193, 0.194 $\mu$m for 
x = p = 0.10, 0.15, 0.20, 0.22 and 0.24, respectively. 
Here p is the hole content per planar copper atom. The estimated 
error for $\lambda_{ab}(0)$ obtained by the $acs$ technique is $\pm
15\%$ and within this uncertainty the $\lambda_{ab}(0)$ values agree
with the $\mu$SR measurements. We thus find that
$\lambda_{ab}^{-2}(0)$ is strongly suppressed on the underdoped side, 
including optimal doping, but there is no suppression with increasing 
overdoping (up to p = 0.24) in contrast to reports for 
Tl$_2$Ba$_2$CuO$_{6+\delta }$ [14]. Values of $\lambda_{ab}(0)$ as obtained
by $\mu$SR for Hg-1201 are 0.194 and 0.148 $\mu$m for $\delta$ = 0.10 and
0.37, respectively. We note that $\delta$ = 0.10 and 0.37 in Hg-1201 
correspond to p = 0.075 and 0.22, respectively [15].
 
The T-dependence of $\lambda_{ab}$ for LSCO is shown in
Fig. 1(a) as a plot of $[1/\lambda_{ab}(T)]^2 \propto \rho_{s}(T)$. 
Data for x =  0.10 and 0.15 obtained by $\mu$SR are also included for 
comparison. Overall there is good agreement between the results from
the two techniques. From the $acs$ data we find that the initial
linear term in $\lambda_{ab}(T)$, characteristic of a clean d-wave 
superconductor, persists up to the highest doping measured (x = 0.24)
in agreement with electronic specific heat studies on polycrystalline 
LSCO samples from the same batch as those studied here [16]. 
Figure 1(b) depicts data for Hg-1201 powders measured only by $\mu$SR,
including data from Ref. [17] for a Hg-1201 sample (also from Houston)
with $\delta$ = 0.154 (p = 0.17). As in LSCO, we observe a change in the
shape of $\sigma (T) \propto$ $[1/\lambda_{ab}(T)]^2$ of Hg-1201 
with doping. Namely, in the underdoped region $[1/\lambda_{ab}(T)]^2$ 
shows a more pronounced curvature. Taking the slope of the low-T
linear term to be proportional to $\rho_s(0)/\Delta_0$ the observed 
trend of $[1/\lambda_{ab}(T)]^2$ with p would imply that $\Delta_0$ 
remains approximately constant in the underdoped region and 
decreases rapidly with overdoping.

Figure 2 shows a comparison of the present results for LSCO
with specific heat data taken on the same samples [16] where
$\Delta_N$ was observed for x = p $<$ 0.19. In the inset we observe a
good correlation between $[1/\lambda_{ab}(0)]^2$ and [S/T(T$_c$) -
S/T(2K)] both plotted versus Sr content x, where S(T) is the
electronic entropy obtained by integrating the electronic specific
heat coefficient $\gamma$(T) from 0 to T. The quantity [S/T(T$_c$) -
S/T(2K)] is a measure of the energy-dependent NS electronic density of
states (DOS), $g_n$(E), averaged over $\pm 2k_{B}T_c$ around the 
Fermi energy E$_F$. The effect of an energy -dependent DOS on the 
London penetration depth $\lambda_L$ or $\rho_s(0)$ is not usually 
considered in standard theory which implicitly assumes a constant DOS 
and a parabolic E($\vec{k}$) dispersion relation. It has been argued 
elsewhere [5] that 
$\rho_{s}(0) = 4{\pi}^2 \langle {v_x}^2 g_{n}(E) \rangle/e^2$ 
where the average is over an (anisotropic) energy shell $E_F \pm
\Delta_0$. Note that this result agrees with the standard expression
for the NS conductivity and the usual relation between
$\lambda_{L}(0)$ and the real part of the frequency-dependent
electronic conductivity in the normal and superconducting states 
${\sigma_1}^{n}(\omega)$ and $\sigma_1^{s}(\omega)$, respectively. Namely,
$\lambda_L(0)$ is determined by the area under the 
[${\sigma_1}^n(\omega) - {\sigma_1}^s(\omega)$] curve in the frequency 
range $0 < h\omega/{2\pi} < 2 \Delta_0$. Thus the inset to Fig. 2 
suggests that the strong decrease of $\rho_s(0)$ with x from 
x = 0.20 to 0.10 is related to the suppression of spectral weight with
energy range $E_F \pm \Delta_0$ due to the presence of $\Delta_N$.
 
The main panel in Fig. 2 shows a correlation between the
doping dependence of the initial linear terms of $\lambda_{ab}(T)$ and
the low-T specific heat coefficient $\gamma$, both quantities being 
related to the number of excited quasi-particles n$_e$(T). 
For low values of x, n$_e$(T=10) is much smaller than expected from 
the T$_c$ value and this probably implies that the average value of 
$\Delta_0(\phi)$ is significantly larger than T$_c$. 
The rapid rise above x = 0.20 may arise from the combined effects of 
the closure of $\Delta_N$ at x = 0.19 [4,16], the decreasing T$_c$
plus the fact that for LSCO there is significant pile up of states
near E$_F$ in the overdoped region $0.20 < x < 0.35$ [16].

In Fig. 3(a) we present the LSCO $acs$ data as 
$[\lambda_{ab}(0)/\lambda_{ab}(T)]^2$ versus T/T$_c$ and compare the 
data with the mean-field calculation for a d-wave weak-coupled BCS 
superconductor with a cylindrical Fermi surface (FS) which gives 
$\Delta_0/T_c \sim 2.14$ [18]. There appears to be a systematic
deviation of the data from the weak-coupling T-dependence with a 
greater (weaker) curvature on the underdoped (overdoped) side. We note
however, that particularly in the overdoped samples there is positive 
curvature near T$_c$ which may arise from a small amount of doping 
inhomogeneity giving a distribution of T$_c$ values in this region
where $dT_c/dp$ is maximal [10]. The effect of this is to rescale the 
curves with a slightly lower value of T$_c$. We have modelled
$\rho_s(T)$ using the d-wave T-dependence and a normal distribution 
of T$_c$ values with standard deviation of 3\%, 5\% and 9\% 
for x = 0.20, 0.22 and 0.24, respectively. 
These corrections, plotted in Fig. 3(b), bring the curves 
for x = 0.20 and 0.22 into good agreement with weak-coupling BCS with 
$\Delta_0/T_c \sim 2.14$. Similar deductions, as to the magnitude and 
p-dependence of $\Delta_0/T_c$, were made in the specific heat studies
on these overdoped samples [16]. However, the x = 0.24 sample still 
shows significant deviations that possibly reflect changes in the 
electronic structure. This would not be surprising given the changes
in the FS with the rapid crossover from hole-like to electron-like
states near x = 0.27 [19]. We note that the data for x = 0.24 is in 
excellent agreement with a weak-coupling d-wave calculation for a 
rectangular FS [20].
 
In contrast to the overdoped samples the optimal and
underdoped samples [Fig. 3(a)], both possessing very small rounding
near T$_c$, diverge significantly from the weak-coupling curve and in
the opposite direction. We note that accounting for inhomogeneities in
these samples will, if anything, move the curves even further from the
weak-coupling BCS fit.
 
A central conclusion of the present work is that there is a 
crossover in both $\rho_s(0)$ and $\rho_s(T)$ near p = 0.20. 
Such behaviour is characteristic of many other NS and superconducting 
properties which have been interpreted in terms of the presence of 
$\Delta_N$ in the underdoped region. The rate of depression of T$_c$
due to impurity scattering ($ \propto 1/\gamma$ at T$_c$) remains 
constant across the overdoped region then rises sharply with the
opening of $\Delta_N$, beginning in the lightly overdoped region at 
p $\sim$ 0.19 [21]. Boebinger and coworkers [22] using intense pulsed 
magnetic fields observe a crossover from insulating to metallic
behaviour at T = 0 occurring near p = 0.18 and angle-resolved
photoemission studies show the development, in the overdoped region,
of a full NS Fermi surface [23]. In this region the resistivity
coefficient, $[\rho (T) -\rho (0)]/T$, exhibits a low-T suppression due
to the opening of $\Delta_N$ [24]. These considerations provide a 
compelling motivation for interpreting our penetration depth data
within a similar scenario.

The proper means of incorporating the pseudogap effects within
a realistic model, and indeed the very nature of the pseudogap is a
matter of current debate. However, a key characteristic of $\Delta_N$
is the loss of NS spectral weight near E$_F$. The loss of spectral
weight can cause, as discussed above, both a strong reduction in 
$\rho_s(0)$ and, in a simple model, enhanced curvature in 
$\rho_s (T)/\rho_s (0)$ above the BCS weak-coupling d-wave
T-dependence [25], the very features we observe for the optimal 
and underdoped samples.
 
We note that our data are in reasonable agreement with earlier
reports for slightly underdoped grain-aligned
HgBa$_2$Ca$_2$Cu$_3$O$_{8+\delta }$ [11,26] and single crystal LSCO with 
x = 0.15 [27]. In contrast to the strong p dependence we have found in
$[\lambda_{ab}(0)/\lambda_{ab}(T)]^2$ for LSCO and Hg-1201, 
studies in YBCO [28,29] reported that
$[\lambda_{ab}(0)/\lambda_{ab}(T)]^2$ scaled approximately with
$1/T_c$ for various dopings, at all temperatures. However, systematic 
changes in $[\lambda_{ab}(0)/\lambda_{ab}(T)]^2$ with p were noted at 
the time [29] although these were too small to allow further
analysis. This may simply be due to the fact that the YBCO samples
were not as heavily underdoped as the x = 0.10 LSCO sample. 
We also note that YBCO is complicated by a mixed s+d order parameter
[1,30] and the effect of the Cu-O chains on the total $\rho_s$ [31,32].  
	
In summary, using the $acs$ and $\mu$SR techniques we have 
obtained consistent and systematic results on the effects of carrier 
concentration on $\rho_s$ of monolayer cuprates. In the overdoped
region we find a more or less constant value of $\rho_s(0)$ (up to p =
0.24), and $\rho_s(T)/\rho_s(0)$ is in reasonably good agreement with
the weak-coupling d-wave T-dependence. In the optimal and underdoped 
regions $\rho_s(0)$ is rapidly suppressed and above 0.1T$_c$ there 
is a marked departure of $\rho_s(T)/\rho_s(0)$ from the weak-coupling 
curve. In a comparative study with available specific heat data we 
found evidence supporting a link in the behaviour of $\rho_s$ and the 
normal state gap $\Delta_N$.
	
We thank P.A. Lee, P.B. Littlewood, T. Xiang and J.F. Annett
for stimulating discussions; J. Chrosch for assistance with part of
the x-ray analysis of the grain-aligned samples, and P. King and
C. Scott (ISIS) for technical support during the $\mu$SR
measurements. C.P. thanks Trinity College, Cambridge for financial support 
and J.B. the European Union for a Marie Curie grant.

\end{multicols}

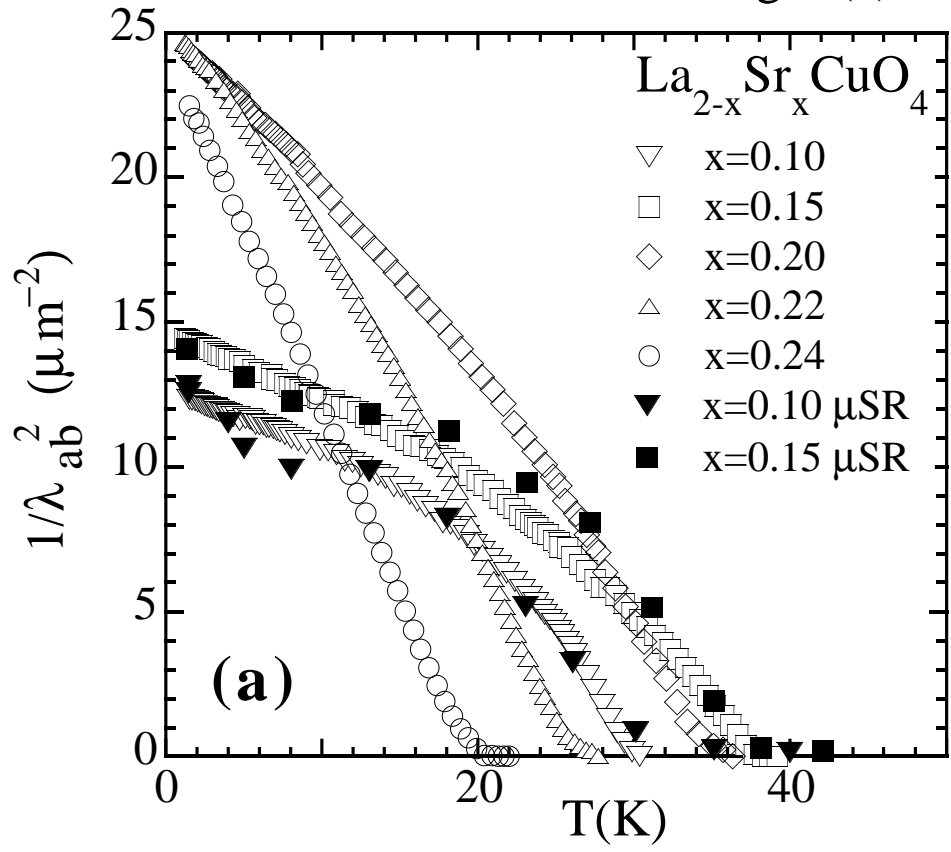
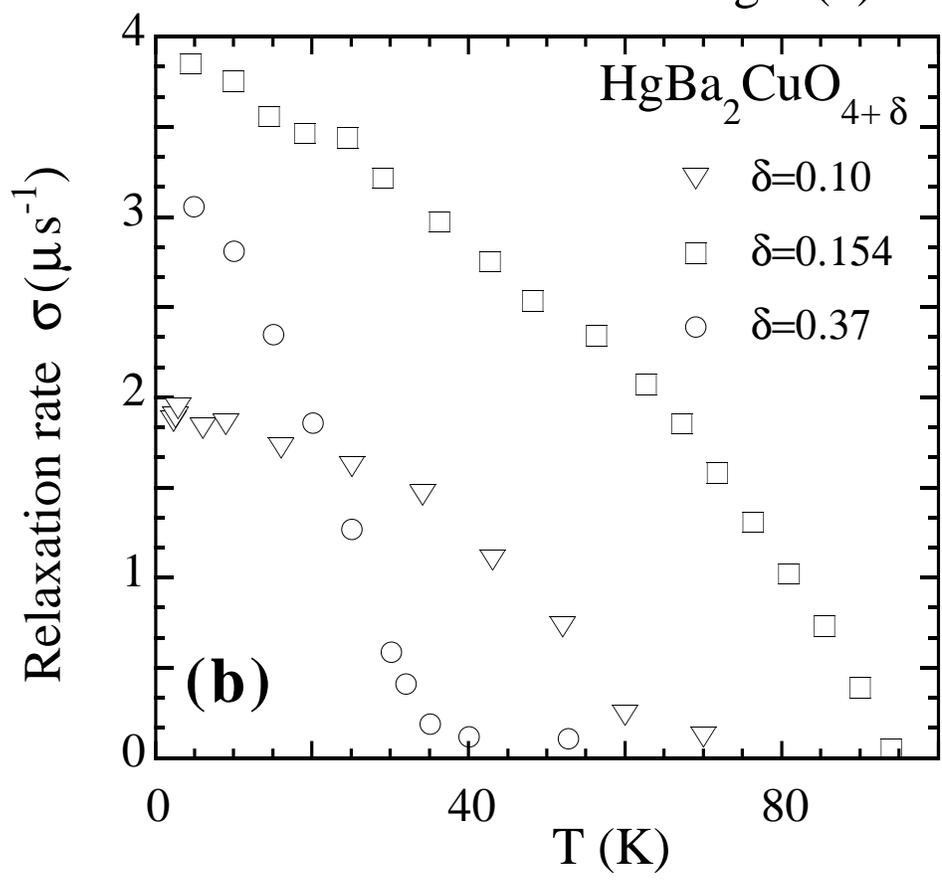
\begin{figure}
\caption{
(a) ${\lambda_{ab}}^{-2}(T)$ obtained by the ac-susceptibility 
technique for grain-aligned La$_{2-x}$Sr$_x$CuO$_4$ (LSCO) with x = p
= 0.10, 0.15, 0.20, 0.22 and 0.24. Data obtained by $\mu$SR for
unoriented LSCO (p = 0.10, 0.15) powders are also included 
(closed symbols). (b) $\sigma (T) \propto \lambda_{ab}^{-2}(T)$, 
for HgBa$_2$CuO$_{4+\delta }$ unoriented powders with 
$\delta$ = 0.10, 0.154 and 0.37 (p  = 0.075, 0.17 and 0.22,
respectively). The data for $\delta$ = 0.154 is taken from Ref. [17].}
\label{FIG. 1}
\end{figure}

\begin{figure}
\caption{
Low-T ${\lambda_{ab}}^{-2}(T)$ for La$_{2-x}$Sr$_x$CuO$_4$ 
(LSCO) versus x compared with the low-T specific heat coefficient 
$\gamma$ [16]. Inset: ${\lambda_{ab}}^{-2}(0)$ for LSCO compared 
with [S/T(Tc) - S/T(2K)] [16].}
\label{FIG. 2}
\end{figure}

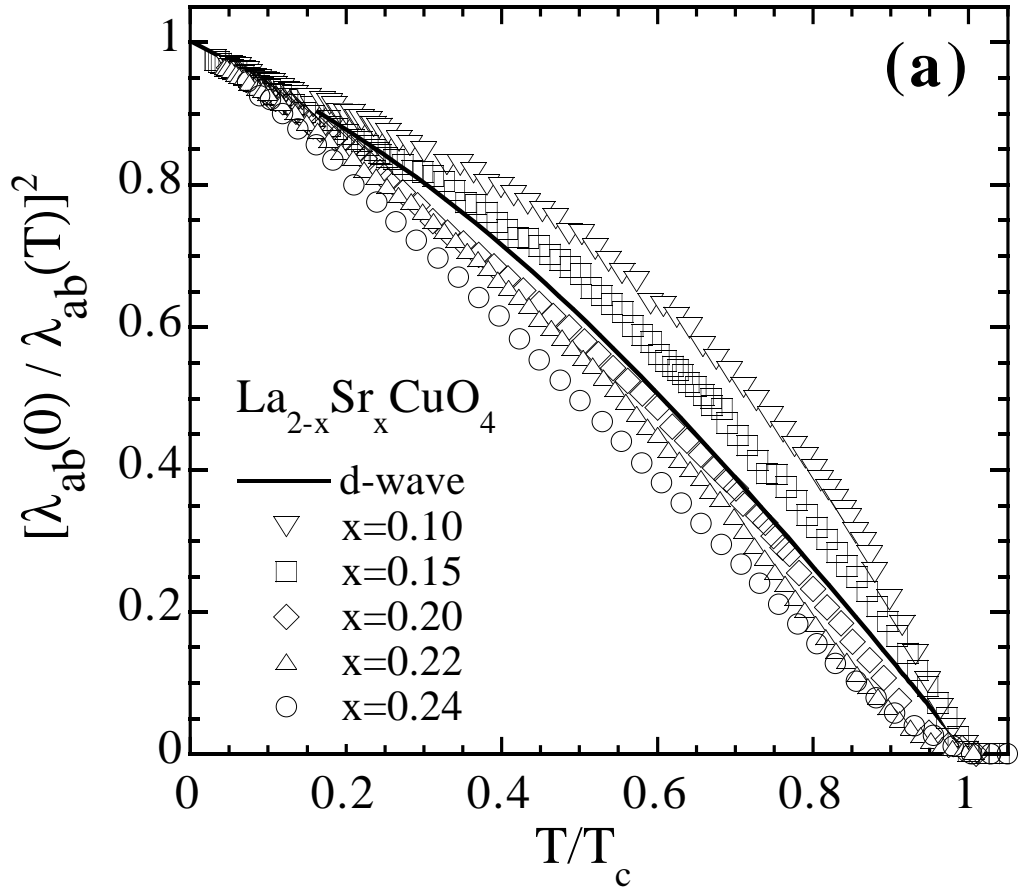
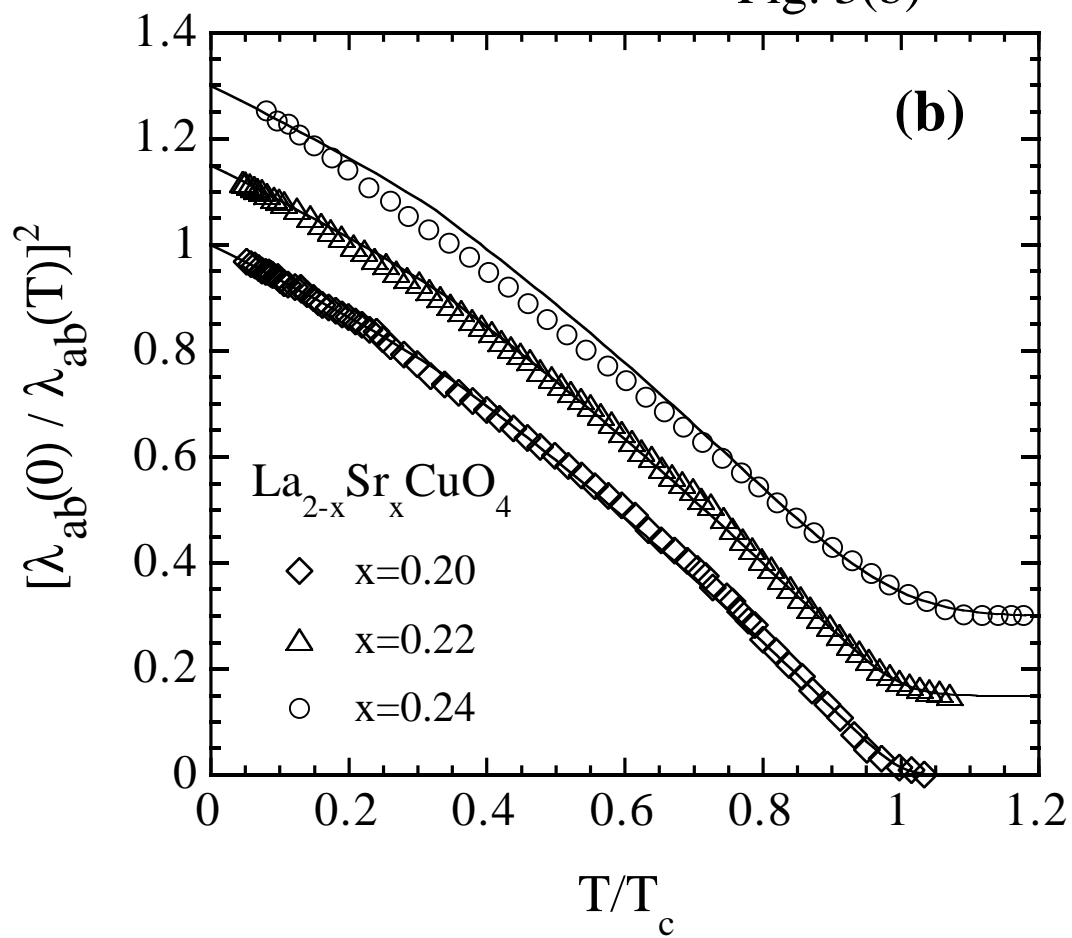
\begin{figure}
\caption{
(a) $[\lambda_{ab}(0)/\lambda_{ab}(T)]^2$ obtained by the 
ac-susceptibility technique for grain-aligned La$_{2-x}$Sr$_x$CuO$_4$ with 
x = p = 0.10, 0.15, 0.20, 0.22 and 0.24 compared with the weak-coupling 
BCS theory (solid line) for a d-wave superconductor [18]. 
(b) The $[\lambda_{ab}(0)/\lambda_{ab}(T)]^2$ 
 data for x = 0.20, 0.22 and 0.24 shown in
panel (a) but corrected for a distribution of T$_c$ values with
standard deviation 3\%, 5\% and 9\% respectively (see text for details). 
The solid lines are the BCS d-wave T-dependence corrected for the 
respective distribution in T$_c$'s. The curves for x = 0.22 and 0.24
are shifted vertically for clarity.}
\label{FIG. 3}
\end{figure}

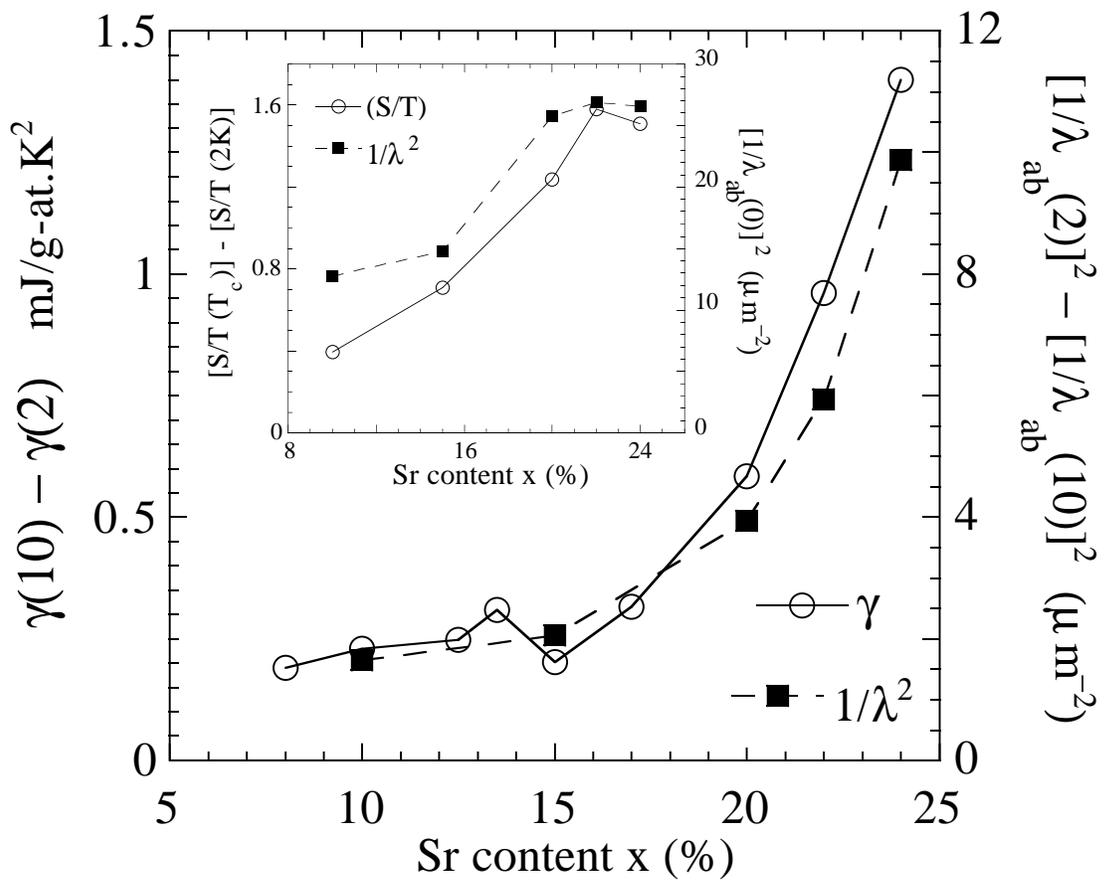
\end{document}